\documentstyle[prl,aps,epsf,floats]{revtex}
\begin{document}
\twocolumn[\hsize\textwidth\columnwidth\hsize\csname@twocolumnfalse%
\endcsname
\draft
%

\title{\bf Photoemission Studies on 
Bi$_{2}$¥Sr$_{2}$¥Ca(Cu$_{1-x}$¥Zn$_{x}$¥)$_{2}$¥O$_{8+\delta}$¥ - 
Electronic Structure Evolution and Temperature Dependence}

\author{P.J.~White$^{(1)}$, Z.-X.~Shen$^{(1)}$, D.L.~Feng$^{(1)}$, 
C.~Kim$^{(1)}$, M.-Z.~Hasan$^{(1)}$, J.M.~Harris$^{(1)}$, 
A.G.~Loeser$^{(1)}$, H.~Ikeda$^{(2)}$, R.~Yoshizaki$^{(2)}$, 
G.D.~Gu$^{(3)}$, N.~Koshizuka$^{(4)}$}

\address{$^{(1)}$Department of Applied Physics and Stanford 
Synchrotron Radiation Laboratory, Stanford University,
Stanford, CA 94305-4045}

\address{$^{(2)}$ Institute of Applied Physics and Cryogenics Center, 
University of Tsukuba, Tsukuba, Ibaraki 305 Japan}

\address{$^{(3)}$ School of Physics, The University of New South 
Wales, PO Box 1, Kensington, N.S.W., Australia 2033}

\address{$^{(4)}$ Superconductivity Research Laboratory, ISTEC, 10-13 
Shinonome, 1-Chome, Koto-Ku, Tokyo 135, Japan}

\maketitle


\begin{abstract}

An angle resolved photoelectron spectroscopy study was conducted on 
Bi$_{2}$¥Sr$_{2}$¥Ca(Cu$_{1-x}$¥Zn$_{x}$¥)$_{2}$¥O$_{8 + \delta}$¥.  A 
small amount of Zn substitution for Cu almost completely suppresses 
the otherwise sharp spectral peak along the (0,0) to $(\pi,\pi)$ 
direction in Bi$_{2}$¥Sr$_{2}$¥Ca(Cu$_{1-x}$¥Zn$_{x}$¥)$_{2}$¥O$_{8 + 
\delta}$¥, while superconductivity with T$_{c}$¥ as high as 83K 
survives.  This behavior contrasts markedly from that seen in cases 
where the impurities are located off the CuO$_{2}$¥ plane, as well as 
when the CuO$_{2}$¥ planes are underdoped.  This effect is also 
accompanied by changes of low energy excitations at $(\pi,0)$, near 
the anti-node position of the d$_{x^{2}¥-y^{2}¥}$¥ pairing state.  
With Zn doping the size of the superconducting gap is significantly 
suppressed, the width of the quasiparticle peak in the superconducting 
state becomes wider, and the dip at higher binding energy is 
diminished.  In addition, enhanced temperature induced spectral changes 
also occur.  We show intriguing systematic lineshape changes with 
temperature that persist to a very high energy scale - a result 
consistent with the idea that Zn enhances the local charge 
inhomogeneity.

\end{abstract}

]
\narrowtext
\section{Introduction}

Impurity doping has been a very effective tool to probe the properties 
of cuprate superconductors.  In particular, Zn substitution of Cu in 
the CuO$_{2}$¥ planes is known to suppress T$_{c}$¥~\cite{Maeda} - 
\cite{Kluge}.  An extensive amount of experiments has been conducted 
to understand the effect of Zn doping, including specific heat, 
microwave, NMR, $\mu$SR, optical, neutron, and tunnelling experiments 
\cite{Hancotte} -\cite{Yamada}.  Transport and microwave experiments 
indicate that Zn is a very strong scatterer, resulting in a strong 
increase in the residual resistivity in the plane~\cite{Chien} 
\cite{Fukuzumi} \cite{Bonn}.  A similar conclusion was drawn from 
$\mu$SR experiments, which also found that Zn suppresses the 
superfluid density~\cite{Bernhard}.  Specific heat, transport, and 
optical experiments indicate that Zn doping alters the residual density of 
states and affects the low energy charge dynamics~\cite{Loram} 
-\cite{Mizuhashi} \cite{Wang}. NMR and neutron experiments generally 
show that Zn introduces low lying excitations in the spin 
channel~\cite{Hirota} -\cite{Tranquada} and dramatically 
affects the dynamical spin fluctuations.  In particular, the NMR 
experiments show that Zn induces local magnetic moments in the normal 
state that do not otherwise show local moment behavior.

More recently, Zn doping was also found to enhance the T$_{c}$¥ 
suppression and other anomalies near $\frac{1}{8}$ doping in 
La$_{2-x}$¥Sr$_{x}$¥CuO$_{4}$¥ (LSCO), YBa$_{2}$¥Cu$_{3}$¥O$_{7}$¥ 
(YBCO), and Y doped Bi$_{2}$¥Sr$_{2}¥$CaCu$_{2}$¥O$_{8+\delta}$¥ 
(Bi2212) \cite{Koike}.  This latter result has been speculated to be 
due to the charge stripe instability \cite{Koike}, similar to a 
possible interpretation of neutron data from LSCO. In addition, a 
neutron scattering experiment has shown two aspects of Zn doping.  
First, it found that Zn shortened the correlation length of the static 
spin density wave \cite{Yamada}.  Second, it found that the Zn shifted 
the spectral weight to lower energy, a fact consistent with the idea 
that Zn serves to stabilize a short range order incommensurate spin 
density wave state which might otherwise be purely dynamic 
\cite{Hirota} \cite{Yamada}.

This issue of microscopic phase separation is of current interest.  
Based on incommensurate neutron scattering data from LSCO (recently 
also observed in YBCO \cite{Mook}), it has been proposed that the 
cuprates develop stripes at low temperature in certain doping regimes 
\cite{Tranquada} \cite{Yamada}.  Here the stripes refer to 
microscopically phase separated insulating and metallic regions 
forming spin and charge ordered one-dimensional structures 
\cite{Emery} - \cite{WhiteScal}.  While the interpretation of neutron 
data is plausible, the experimental evidence for charge ordering 
remains elusive at present.  Except for the case of Nd doped LSCO, no 
evidence of charge ordering has been detected.  Zn doped Bi2212 may be 
a good system for an angle resolved photoelectron spectroscopy (ARPES) 
investigation of this issue.  First, Zn doping is found to enhance the 
T$_{c}$¥ suppression and other anomalies near $\frac{1}{8}$ doping in 
LSCO \cite{Koike}, YBCO \cite{Mook}, and Y doped Bi2212 
\cite{Akoshima}.  The $\frac{1}{8}$ anomaly is thought to be 
associated with the stripe instability \cite{Tranquada}.  Second, from 
what we will show later, Zn impurities dramatically alter the 
electronic structure along the $(0,0)$ to $(\pi,\pi)$ line.  This data 
can be rationalized by the fact that Zn induces an antiferromagnetic 
region around it and enhances the local charge inhomogeneity.

By using ARPES, we will explore these aspects in detail.  Over the 
last decade ARPES has played an important role in advancing our 
understanding of the low energy single particle excitations in these 
novel superconductors, starting with the observation of band like 
features \cite{Olson}.  Most notably, ARPES facilitated the 
observation of the d-wave superconducting gap structure as well as the 
normal state pseudogap \cite{Shen} \cite{Marshall} \cite{Loeser} 
\cite{Ding}.  Several groups have attempted to study the impurity 
doping effects of the electronic structure using ARPES. Quitmann {\it 
et al.\/} \cite{Quitmann} have performed room temperature ARPES of Ni 
and Co doped Bi2212 to address the changes of the electronic structure 
in the normal state but not the changes in the superconducting state.  
Our experiments on Bi2212 will address both the normal and 
superconducting states.  Gu {\it et al.\/} \cite{Gu} have attempted to 
study Zn and Co doped YBCO for which the superconducting property of 
the CuO$_{2}$¥ planes is complicated by the surface chain signal 
\cite{Schabel}; Bi2212 lacks the surface chain complication of YBCO.

\section{Experimental} 
Single crystals of 
Bi$_{2}$¥Sr$_{2}$¥Ca(Cu$_{1-x}$¥Zn$_{x}$¥)$_{2}$¥O$_{8 + \delta}$¥ (Zn 
doped Bi2212, x=0.006,0.01) were prepared using a traveling solvent 
floating zone method \cite{Yoshizaki} \cite{GendaGu}.  These crystals 
were characterized according to T$_{c}$¥ and the Zn concentration, 
which was determined by electron probe microanalysis (EPMA).  The 
crystals were grown under the nominal condition to produce optimal 
doping, with T$_{c}$¥ of the Zn doped samples ranging from 83K to 78K. 
The transition widths vary from 3-5K according to susceptibility 
measurements, indicating the high quality of these crystals.  The 
single phase of the samples was verified by X-ray scattering.  X-ray 
rocking curves indicate that the crystalline quality of the Zn doped 
Bi2212 is comparable to that of the pure Bi2212; the presence of 
stacking faults was checked by taking the rocking curves of the 
(0,0,$\ell$) X-ray diffraction peaks and the results were comparable 
between pure Bi2212 and our Zn doped Bi2212 \cite{Yoshizaki}.  In 
addition, Laue back scattered X-ray diffraction was done for alignment 
purposes and no difference was detected between the standard and the 
Zn doped sample.  In this paper, data from 3 samples of pure Bi2212 
are presented with T$_{c}$¥$\approx$91K, 89K and 88K. The former two 
reveal themselves to be typical samples of high quality, but the last 
one with a critical temperature of 88K exhibits behavior between the 
ones closer to optimal doping and the ones doped with Zn.  Therefore, 
we conjectured that the 88K sample had some unknown impurities.  
Attempts were made to determine the stoichiometry more precisely via 
EPMA, but owing to the number of species in the compound and the low 
concentration of the impurity, this was difficult with conventional 
means.  While this may seem strange to include, it is nevertheless 
presented to make a connection to our previously published results 
\cite{Shen2} and to provide continuity of our work.  This paper is 
also an homage to the previously neglected subtleties on our part 
associated with this kind of doping in Bi2212.  Therefore, we 
elucidate our old results in light of our new findings.

Data in Figs.~\ref{Fig3ver5}, \ref{TonyMbarver3}, 
\ref{Fig5ver5}, \ref{Fig6ver1}, \ref{Znnofk} were recorded with a 
Vacuum Science Workshop analyzer attached to beamline 5-3 of the Stanford Synchrotron Radiation 
Laboratory (SSRL).  The total energy resolution was typically 35meV and the 
angular resolution was $\pm$1$^{\circ}$¥.  The nominal chamber 
pressure during the measurement was 2-3x10$^{-11}$¥ torr and the 
photon energy used was 22.4eV. At this photon energy an ARPES spectrum 
mimics the spectral function, A({\bf k},$\omega$) \cite{Randeria}, 
weighted by the appropriate factors, such as matrix elements, the 
Fermi function, etc.  The ARPES spectra in the remaining data were recorded 
with a Scienta analyzer attached to beamline 10 of the Advanced Light 
Source (ALS).  The total energy resolution was typically 15meV and the 
angular resolution was $\pm$0.15$^{\circ}$¥ with the spectrometer 
operating in angle mode.  The nominal chamber pressure was 
6x10$^{-11}$¥ torr and the photon energy used was 25eV. Spectra from 
SSRL were taken within 10-12 hours of cleaving so as to minimize aging 
effects as previously reported \cite{Shen} \cite{Ding2}.  Spectra from 
ALS were taken within a shorter time to compensate for additional 
aging caused by the higher photon flux.  With the SSRL apparatus, we 
can only take selected {\bf k} points in order to have spectra with 
low enough statistical noise to identify subtle changes in the 
lineshape.  With the ALS apparatus we can take about 40 spectra at the 
same time with 0.15$^{\circ}$¥-0.3$^{\circ}$¥ spacing.  The flatness 
of the surfaces of the pure and Zn doped samples was verified by laser 
reflection patterns after the samples were cleaved {\it in situ}.  
Fermi levels were determined by a gold reference sample in electrical 
contact to the samples.

\section{Electronic Structure Evolution}
In this section we report detailed results of ARPES on the nature of 
the Zn doping effect on the electronic structure of the Zn doped 
Bi2212 system.  We found significant changes in the electronic 
structure near the Fermi level with a small amount of Zn doping.  
Along the (0,0) to $(\pi,\pi)$ direction, Zn doping essentially wipes 
out the otherwise well defined spectral peak \cite{comment} in samples 
with T$_{c}$¥ as high as 83K. This behavior contrasts strongly to the 
case where scattering impurities are located off the CuO$_{2}$¥ plane 
as well as to the case of an underdoped CuO$_{2}$¥ plane.  Zn doping 
also causes systematic changes in data near $(\pi,0)$, which is 
close to the anti-node region of the d-wave pairing state.  Indeed, 
the superconducting gap is decreased as one would expect from pair 
breaking considerations.  At the same time, the dip seems almost gone 
in Zn doped Bi2212.  This suggests an interesting evolution of the 
$(\pi,0)$ superconducting spectrum as the traditionally two distinct 
features (the broad incoherent peak and the sharp spectral peak) seem 
to evolve simultaneously with Zn doping.

\subsection{Experimental Observation} 
\label{Experimental Observation}
Fig.~\ref{Fig3ver5} presents ARPES data at selected {\bf k}-space 
points along (0,0) to $(\pi,\pi)$ for pure, Zn doped, Dy doped and 
underdoped samples.  Two sets of data from the pure and the Zn doped 
samples are shown to illustrate the reproducibility.  These points 
were chosen for their proximity to the Fermi surface and were spaced 
sufficiently in momentum to reveal the behavior of the Fermi level 
crossing.  For data from pure sample in Fig.~\ref{Fig3ver5}a-b, we 
see a relatively sharp feature disperse across E$_{f}$¥ in the 
expected way.  As the peak gets closer to the Fermi level, it 
apparently narrows in width and, at some point, loses intensity until 
it ultimately vanishes.  This observation is consistent with previous 
work \cite{Olson} \cite{Marshall} \cite{Loeser} \cite{Ding}, and much 
of the peak width is attributable to angular and energy resolutions.  
This general behavior is qualitatively what one expects of a 
quasiparticle.  For data from the Zn doped samples from 
Fig.~\ref{Fig3ver5}c-d, the dramatic difference is readily apparent.  
The spectral peak is wiped out with no sharp feature seen at the 
expected crossing or before it \cite{comment2}.  (We note in 
Fig.~\ref{Fig3ver5}d that the peak is not recovered upon cooling.)  
For comparison, Fig.~\ref{Fig3ver5}e-f reproduces our results from two 
underdoped samples in similar {\bf k}-space locations \cite{Marshall}.  
The underdoping in these samples (T$_{c}$¥ near 65K for both) was 
achieved either by removing oxygen or by substituting 10\% Ca by Dy.  
In both cases the feature along $\Gamma$Y remains fairly sharp; this 
contrasts strongly with data from the Zn doped samples in 
Fig.~\ref{Fig3ver5}c-d.  For the 10\%
Dy doped sample there is the additional effect of scattering by Dy 
impurities, whose concentration is much higher than that of the Zn 
impurities.  It is clear that the Zn impurities in the CuO$_{2}$¥ 
plane did far more damage to the quasiparticle peak than the more 
highly concentrated Dy impurities, which are located in the Ca plane 
sandwiched by the CuO$_{2}$¥ planes.

Fig.~\ref{TonyMbarver3} shows the $(\pi,0)$ spectra of the pure and 
the Zn doped samples and reproduces previously published results on 
overdoped and underdoped samples for comparison.  Unlike the $\Gamma$Y 
line as shown in Fig.~\ref{Fig3ver5}, the normal state spectra of the 
pure and Zn doped samples are similar in this region of {\bf k}-space.  
Both samples show very sharp peaks in the superconducting state.  The 
fact that one can see such a sharp peak below T$_{c}$¥ in Zn doped 
samples gave us confidence on the intrinsic nature of the very broad 
feature in Fig.~\ref{Fig3ver5}c-d.  It is possible that a disordered 
surface can produce the effect seen in Fig.~\ref{Fig3ver5}c-d, 
however the coexistence of a disordered surface and the sharp feature 
seen in the data at $(\pi,0)$ is unlikely.

There are several subtle but important differences between the 
$(\pi,0)$ superconducting spectra for the pure and Zn doped samples.  
Readily apparent  is the fact that there is virtually no dip at the higher 
binding energy side of the sharp peak in the Zn doped samples, whereas 
the dip is clearly visible in the pure sample as found before 
\cite{Dessau} \cite{Hwu}.  The superconducting quasiparticle peak is 
also broader in the Zn doped sample, implying a stronger scattering 
rate.  In the Zn free sample, the peak width is resolution limited.  
While the spectral weight of the Zn free sample is balanced above and 
below T$_{c}$¥, the spectral weight of the Zn doped sample is 
increased at lower temperature because of the sharp peak's 
development.  Two changes in the spectra contribute to this 
imbalance: the dip no longer exists and the peak is significantly 
broadened, even though it has the same relative maximum.  The sum 
rule of A$({\bf k},\omega)$ requires spectral weight to come from 
other locations in energy and/or momentum space.  We will address this 
point later.

Another significant observation in Fig.~\ref{TonyMbarver3}a is that 
the sharp peak in the Zn doped sample shifts to lower binding energy 
as compared to that of the Zn free sample.  This can be interpreted as 
the size of the superconducting gap being suppressed in the Zn doped 
sample.  In the literature, the size of the energy gap in 
photoemission is often characterized by the position of the leading 
edge midpoint in the spectra recorded at the underlying Fermi surface 
\cite{Shen} \cite{Marshall} \cite{Loeser} \cite{Ding}.  In this case, 
because the peak is appreciably broadened in the Zn doped sample, the 
leading edge analysis is not ideal.  We use the quasiparticle peak 
position as a way to characterize the gap.  Here, we use the $(\pi,0)$ 
peak even though it is not exactly at the Fermi crossing.  However, 
since the band dispersion is very flat in this region, we can use it 
to track the relative change in the gap size.

The relative changes of the peak in Fig.~\ref{TonyMbarver3}a suggest 
the superconducting gap is suppressed in the Zn doped samples.  Note 
the difference in the rate of $\Delta$ suppression and T$_{c}$¥ 
suppression as a function of Zn doping.  It is clear that the gap is 
severely suppressed in the Zn doped samples, given the modest T$_{c}$¥ 
decrease at this doping level.  This was consistent with an earlier 
conclusion that T$_{c}$¥ and gap are not directly related energy 
scales \cite{Emery}.  A possible scenario is that T$_{c}$¥ is not 
limited by pairing strength but by phase fluctuation effects 
\cite{Emery2} \cite{Doniach} \cite{de Melo}.  It is also worth noting 
that the normal state spectrum at $(\pi,0)$ of the Zn doped sample is 
cut off by the Fermi function, ruling out the existence of the normal 
state pseudogap.  This is not the case for Zn free Bi2212 in 
Fig.~\ref{TonyMbarver3}a.  This finding can be interpreted as Zn 
doping suppressing the pseudogap or creating low lying excitations 
inside the gap as reported by other experiments \cite{Kakurai} 
\cite{Sidis}.

Summarizing data from Figs.~\ref{Fig3ver5} and \ref{TonyMbarver3}, we 
observed correlated changes of the electronic structure as a function 
of Zn doping: the strong suppression of the quasiparticle along the $\Gamma$Y 
line; the suppression of the superconducting gap; the broadening of 
the superconducting peak and the suppression of the dip near 
$(\pi,0)$.

\subsection{Discussion of Electronic Structure Evolution Results}
The experimental data presented raise several interesting points about 
impurity doping in the cuprates.  The first and foremost has to do 
with whether the conduction mechanism can be described by 
quasiparticle dynamics.  The dramatic differences in the ARPES data of 
Fig.~\ref{Fig3ver5}c-d with a small amount of Zn doping is unexpected 
from the doping dependence studies of other materials.  Photoemission 
is a signal averaging experiment and is usually quite insensitive to a 
small amount of doping change, unlike the case here.  In transition 
metal oxides one usually sees only subtle changes with doping 
variation up to 10-20\% \cite{Morikawa}.  In ordinary metals like Cu 
or Al the spectra do not change with a very small amount of 
impurities.  This is not the case for Zn doping; the spectra change 
dramatically.  The fact that the change in the Zn spectra in the normal state was consistent with the 
change in the superconducting state gave us confidence on 
the intrinsic nature of this effect (again we note that the very sharp 
peak below T$_{c}$¥ at $(\pi,0)$ shows it was not due to a 
contaminated surface).  The magnitude of the change with a relatively 
small amount of Zn suggests the system may be very close to a certain instability, and the effect of the Zn impurity is amplified by this 
intrinsic instability.  In Zn free samples the sharp, dispersive 
feature along the $\Gamma$Y direction resembles what one would expect 
from quasiparticles with well defined {\bf k}, although the feature is 
still too broad for this description.  On the other hand, 
Zn doped Bi2212 showed that there is no quasiparticle with well defined {\bf 
k} at all in this direction.  Given the modest change of T$_{c}$¥, the 
change seen in Fig.~\ref{Fig3ver5}c-d is quite remarkable.  It 
suggests that normal state quasiparticles with well defined momenta 
are not essential for superconductivity.  The observed change of the 
low energy electronic structure in Fig.~\ref{Fig3ver5}c-d is 
consistent with reports from other experiments.  NMR, specific heat, 
microwave, optics, and transport experiments indicate that Zn doping 
alters the residual density of states \cite{Loram} -\cite{Mizuhashi} 
\cite{Wang} and affects the low energy and spin dynamics 
\cite{Bernhard} -\cite{Tranquada}.  The {\bf k}-resolved information 
from ARPES is new.

To emphasize the peculiar effect of only a tiny amount of Zn (0.6\%, 
average concentration), 
we wish to examine what one would expect from simple considerations.  
Naively, one would think that Zn doping causes scattering in the 
CuO$_{2}$¥ plane, and this would cause an angular averaging effect; 
this is similar to the scattering of an incident electron with 
wavenumber {\bf k} by an impurity and mixing with the scattered 
spherical wave, which is made of a range of {\bf k}.  To mimic this 
process we averaged the spectra from the six angles in 
Fig.~\ref{Fig3ver5}a and still obtained a reasonably sharp peak 
(uppermost curve of Fig.~\ref{Fig3ver5}a).  As parallel cuts are 
similar in the nearby regions, inclusion of these cuts in the 
averaging process would give a similar effect to what we have done 
here.  The averaged data still had a sharper structure than those in 
Fig.~\ref{Fig3ver5}c-d.  Considering the Zn doping is only 0.6\%, the
scattering effect would be much less than the averaging process as we 
have done here even if Zn acts as a scatterer.  Therefore, the notion 
of Zn being a simple scatterer seems insufficient to explain the 
effect seen in Fig.~\ref{Fig3ver5}c-d.  An alternate hypothesis is 
required.  It may be that the Zn impurities induce some collective 
effects like the ones suggested by neutron experiments.  One 
collective effect is that Zn impurities induce long range 
antiferromagnetic order that co-exists with the spin-Peierls 
transition seen in in Cu$_{1-x}$¥Zn$_{x}$¥GeO$_{3}$¥ \cite{Hase} 
\cite{Manage} \cite{Sekine}.

Another is that the Zn impurities may pin the dynamical stripes 
\cite{Hirota} \cite{Tranquada}.  We will now explore whether this idea 
provides a self consistent explanation to our data.  Specifically, we 
want to see whether the data are compatible with the idea that Zn 
impurities pin the dynamical stripes.  There are two reasons for us to 
consider this possibility.  The first has to do with transport 
experiments at $\frac{1}{8}$ doping \cite{Koike}.  It has been 
strongly suggested that the T$_{c}$¥ suppression and other transport 
anomalies at $\frac{1}{8}$ doping are related to the stripe 
instability \cite{Tranquada}.  The work on Zn-Y doped Bi2212 is 
particularly relevant to our discussion here \cite{Akoshima}.  In Zn 
doped cases (2-3\%), it is found that the electrical
resistivity and thermoelectric power exhibit less metallic behavior 
than usual near a doping level of $\frac{1}{8}$.  At the same time, 
T$_{c}$¥'s for samples near a doping level of $\frac{1}{8}$ are also 
anomalously suppressed.  These results suggest that the Zn doped 
Bi2212 system has certain similarity to the LSCO system where data are 
interpreted as possible evidence that Zn pins the dynamical stripes.  
The second reason concerns results from neutron experiments.  Recent 
neutron scattering data from Zn doped LSCO indicate that Zn doping 
shifts the spectral weight of the incommensurate peaks at 
$(\pi,\pi\pm\delta\pi)$ to lower frequencies \cite{Hirota}.  As the 
incommensurate peaks are interpreted as scattering from dynamical 
stripes \cite{Tranquada}, the downward shift of spectral weight has 
been interpreted as stablization of the dynamic stripes \cite{Hirota}.  
On the other hand, it is also found that Zn broadened the 
incommensurate neutron peak, which indicates that Zn doping disrupts 
the long range order and shortens the correlation length.  Hence, it 
appears that a random distribution of Zn impurities shortens the long 
range correlation but stabilizes the short range correlation which 
would otherwise be more dynamic \cite{Tranquada}.

Now, we wish to draw upon the phenomenological similarities between 
the ARPES data for LSCO and Bi2212 along (0,0) to $(\pi,\pi)$.  
Empirically, ARPES features along (0,0) to $(\pi,\pi)$ are always 
sharp, even in underdoped materials (Fig.~\ref{Fig3ver5}e-f).  This is 
also the case for YBCO and Bi2212 systems \cite{Schabel} \cite{Liu} 
\cite{Harris} \cite{Gofron} as well as the insulating 
Sr$_{2}$¥CuO$_{2}$¥Cl$_{2}$¥ and Ca$_{2}$¥CuO$_{2}$¥Cl$_{2}$¥ 
\cite{Wells}.  The only cuprate that violates this empirical rule is 
the LSCO system.  The data from LSCO system show strong resemblances 
to that of the Zn doped system here \cite{Ino}.  In the LSCO case 
spectra along the (0,0) to $(\pi,\pi)$ direction are found to be 
extremely broad, while one can still see well defined peaks near 
$(\pi,0)$ for highly doped cases - a fact that shows that the 
extremely broad feature along the (0,0) to $(\pi,\pi)$ direction is 
not due to a bad surface.  With the increase of Sr doping, the change 
of the spectra along (0,0) to $(\pi,\pi)$ is not monotonic.  For 
x$<$0.05, one sees a broad dispersive feature that behaves like the 
feature seen in the insulating Sr$_{2}$¥CuO$_{2}$¥Cl$_{2}$¥.  For 
0.05$<$x$<$0.15, one hardly sees any dispersive feature at all.  For 
x$>$0.2 one again sees a dispersive feature that shows a clear Fermi 
level crossing.  Hence, the broadness of the feature along (0,0) to 
$(\pi,\pi)$ is not simply due to the random disorder of the Sr 
impurities as the Sr content increases monotonically.  There is 
another likely contribution to the broadness of the features, and the 
contribution might be connected to the intrinsic electronic 
inhomogeneity.  Based on neutron experiments, the evidence for stripe 
correlation is strongest in the LSCO system and is most visible in the 
doping range near $\frac{1}{8}$ \cite{Tranquada}.  Given the 
similarity of the data from Zn doped Bi2212 and from LSCO, one may 
wonder whether the destruction of the spectral peak along (0,0) to 
$(\pi,\pi)$ in the Zn samples is also related to the electronic 
inhomogeneity.  This issue is also related to the recent transport 
measurement on Zn-Y doped Bi2212 \cite{Akoshima}.

Extensive numerical calculations using Hubbard or t-J models have been 
carried out on small cluster samples \cite{Dagatto}.  The systematics 
of the spectral lineshapes seen in Bi2212, YBCO, 
Sr$_{2}$¥CuO$_{2}$¥Cl$_{2}$¥, Nd$_{2-x}$¥Ce$_{x}$¥CuO$_{4}$¥, as well 
as the doping dependence, can very well be accounted for by these 
calculations.  However, the spectral behavior of Zn doped Bi2212 and LSCO 
cannot be explained by these calculations even qualitatively.  All 
calculations produce a sharp structure along (0,0) to $(\pi,\pi)$, 
even when additional parameters such as t$^{\prime}$¥ and 
t$^{\prime\prime}$¥ are included \cite{Kim2}.  On the other hand, an 
exact diagonalization of the spectral function shows that an 
introduction of attractive forces along the $(\pi,0)$ or $(0,\pi)$ directions 
strongly suppresses the quasiparticle-pole strength along (0,0) to 
$(\pi,\pi)$ direction \cite{Tokyama}.  These forces were introduced to 
simulate the effects of stripes on the electronic structure, although 
the origin of stripes is unclear.  In another calculation on the t-J 
model, it is found that a Zn vacancy pins the domain wall, similar to 
the case where a Zn vacancy bounds a hole of d$_{x^{2}¥-y^{2}¥}$¥ 
symmetry \cite{Poilblanc}.  Summarizing the above discussion, it 
appears that the scenario of electronic phase separation can provide a 
self consistent accounting of the data.  However, the above 
interpretation is not unique, and the phase separated regions do not 
necessarily have 1D order.

Alternatively, the interpretation of the data may come from a more 
phenomenological viewpoint.  Zn does have an effect on the electronic 
properties of the cuprates.  One particular result is that Zn induces 
local moments on neighboring Cu sites \cite{Alloul}\cite{Mahajan}.  
This glassy array of antiferromagnetic droplets will probably become 
more ordered as the temperature is lowered \cite{PatrickLee}.  This 
will affect the electronic structure, and it is possible that the 
phenomena here are just manifestations of that fact.  At the same 
time, quasiparticles at $(\frac{\pi}{2},\frac{\pi}{2})$ may suffer 
significant scattering, resulting in the washed out features we see 
here.  The spectral lineshape changes at $(\pi,0)$ are not surprising 
as Zn has an effect on the superfluid density, n$_{s}$¥.  According to 
Nachumi \/{\it et al.\/} \cite{Nachumi}, the Zn impurity acts as a 
dead center for n$_{s}$¥.  This pocket of no superfluid extends over 
an area of $\pi\xi^{2}$¥, where $\xi$ is the in-plane coherence 
length.  This also says that the volume available for 
superconductivity is reduced, and this ought to be reflected as a 
decrease in n$_{s}$¥.  Comparing the relative strength of the 
$(\pi,0)$ peak with and without Zn, it is hard to say that the data 
points towards this conclusion.  Furthermore, one should compare 
samples with the same $\delta$, and that has been a hard parameter to 
control in the growth of Bi2212.  It should be noted that while we 
have included this for the sake of discussion, the role of Zn in 
forming local moments in underdoped cuprates is still an open issue 
\cite{Williams}.

The next issue concerns the spectral lineshape of the photoemission 
experiments.  The systematic changes in ARPES data with Zn doping 
provide a new perspective on several long standing problems about the 
unusual photoemission lineshape observed in high T$_{c}$¥ 
superconductors.  These problems can best be illustrated by the 
spectral lineshape change at $(\pi,0)$ above and below T$_{c}$¥, as 
shown in Fig.~\ref{TonyMbarver3}.  As the temperature is lowered below 
T$_{c}$¥, a sharp quasiparticle peak emerges at the low energy edge of 
the broad normal state feature accompanied by a dip structure at 
higher energy.  In Zn doped Bi2212 the dip is gone, but the peak 
persists and even gains intensity as it is broadened but with roughly 
similar height.  It is often assumed that the sharp peak develops 
below T$_{c}$¥ because of an increase in quasiparticle lifetime, which 
is independently measured in other experiments \cite{Bonn2} 
\cite{Harris2}.  The traditional interpretation of the peak and dip 
structure is associated with the electronic pairing mechanism 
\cite{Littlewood}.  In this case, the peak is the superconducting 
quasiparticle at gap energy $\Delta$ (for the case when the normal 
state quasiparticle peak is at the Fermi level) and the dip is caused 
by the suppression of the spectral weight between the energy $\Delta$ 
and $3\Delta$ as the electronic medium itself is gapped.  More 
recently, a phenomenological self-energy was proposed to explain the 
peak and the dip in a very similar spirit \cite{Norman2}.  In both of 
these cases, the dip and the peak go hand in hand because they are 
manifestations of the same self-energy change.

The data from Zn doped Bi2212 add to a list of puzzles associated with the 
above interpretation as Zn doping kills the dip without diminishing 
the intensity of the peak so that the peak and dip do not necessarily 
follow each other.  The other puzzles are represented by the 
following examples.  First, the expectation of spectral weight 
suppression between $\Delta$ and $3\Delta$ is based on a very general 
argument compatible with any electronic pairing mechanism so that the 
dip will appear as long as the electronic excitation spectrum opens a 
gap.  This expectation is in strong contrast to the fact that one does 
not see a dip structure in the pseudogap state of the underdoped 
samples where the electronic excitation spectrum opens a gap 
\cite{Loeser}.  It is also very strange that the sharp peak emerges 
only below T$_{c}$¥ in the underdoped samples although the spectrum is 
gapped well above T$_{c}$¥, as if the single particle spectral 
function is sensitive to the superconducting order.  Further, the 
intensity of the sharp peak shows a monotonic correlation with doping 
and, thus, the superfluid density.  These empirical observations are 
not compatible with the conventional wisdom that the single particle 
spectral function should not be sensitive to the superconducting 
condensate.  In fact, we believe this puzzle is a very important clue 
and its understanding will lead to a deeper insight into these 
materials.  Second, the sharp peak and the broad feature at higher 
binding energy side of the dip seem to behave independently 
\cite{comment3}.  Fig.~\ref{TonyMbarver3} shows the $(\pi,0)$ spectra 
in the superconducting state with different doping.  It is clear that 
the energy of the superconducting peak hardly changes, but the broad 
feature and the shape of the dip change significantly.  Third, the 
photoemission spectra obtained in the normal state of the doped 
superconductor show a striking resemblance to that of the insulator, 
albeit the energy position of the broad feature near $(\pi,0)$ evolves 
with different doping.  Finally, we have to consider the fact that 
the peak and dip structures are not seen in YBCO. Despite the chain 
complication, we can still get the consistent result about the Fermi 
surface, the d-wave gap, and the sharp peak at X $(\pi,0)$ in YBCO 
\cite{Schabel}.  However, the dip structure is not seen even in 
measurements where the sample is cleaved and kept at 10K 
continuously.  The above puzzles suggest that we may need to re-evaluate the problem in a 
very different way, as the conventional wisdom cannot explain the data.

\section{Temperature Dependence}
In this section we report the temperature induced ARPES spectral change 
in Zn doped Bi2212.  As with the spectra 
themselves, the temperature induced change varies with doping.  With 
the increase of Zn impurities in the samples, we see a 
strong temperature dependence in the data.  Furthermore, the 
temperature induced spectral change extends to very high energy in Zn 
doped samples.  Empirically, the temperature dependence 
of the overdoped metal is much weaker than that of the undoped 
insulator \cite{Kim}.  In the later case, the temperature 
dependence persists up to an energy scale of about an electron volt.  
Since the carrier density is not changed with Zn substitution of Cu, 
the enhanced temperature dependence is consistent with the idea that 
Zn creates antiferromagnetic regions around it which exhibit stronger 
temperature dependence.

If Zn enhances the tendency of microscopic phase separation, the next 
question is whether the phase separation takes the form of 1D stripes 
as transport experiments seem to suggest \cite{Akoshima}.  The key 
difficulty lies in the fact that the traditional tools ({\it e.g.\/} 
X-ray or electron scattering) are sensitive to all electrons in the 
system, while the charge ordering, if present, occurs for electrons 
close to the Fermi level.  This `large background' signal from other 
electrons makes the detection of any charge density wave very 
difficult.  Angle resolved photoemission may provide an 
opportunity to address this important and current issue as one can 
view ARPES as a kind of photoelectron diffraction experiment.  Since 
the ARPES data near E$_{f}$¥ probe only the last valence electron, it 
is naturally more sensitive to charge inhomogeneities as it 
discriminates against the background signal from other electrons.  
Further, the relatively low kinetic energy of these photoelectrons 
make them more sensitive to any charge density modulation than the 
high energy electrons and X-rays used in typical experiments.  The 
complication, of course, is that these photoemitted valence electrons 
will be multiply scattered by other electrons (including the core 
electron) - a complex process itself is the source of information and, 
thus, cannot be avoided.  This will reduce the sensitivity and 
increase the difficulty in the data analysis.  Nonetheless, ARPES may 
still have higher sensitivity as it has the energy selectivity so one 
can probe specific orbitals. 

Motivated by this consideration, we have made attempts to use ARPES to 
investigate the issue of charge stripes in cuprates.  If charge 
stripes develop at low temperatures, as indirectly suggested by 
neutron scattering experiments, it may diffract the photoelectrons and 
cause a {\bf k}-dependent change in integrated spectral weight, n({\bf 
k}).  In an earlier report \cite{Shen2}, we presented data that 
suggested a {\bf q} dependent spectral weight shift, and we have 
attributed that to the stripe instability because the {\bf Q} value so 
determined is quite close to what one expects from charge stripes.  We have also 
observed the enhanced effect of the {\bf q} dependent spectral weight 
shift in the Zn doped samples.  This latter aspect of the data, 
however, exhibits strong scatter and cannot be made unambiguous.  
Therefore, we have also detailed the uncertainties associated with our 
interpretation of the {\bf q} shift of the spectral weight and its 
connection with charge stripes.  In this context, we have also 
discussed the claim that the spectral weight to be temperature independent 
\cite{Campuzano}.  We disagree with this premise and we show the prior 
data and claims of the same authors actually pointed to the 
contrary.  We believe there are systematic changes of the spectra with 
temperature which depend on the doping and impurity concentration in 
the sample.  On the other hand, the connection between the temperature 
dependence in the ARPES data and the stripe picture is not as 
straightforward as we have suggested \cite{Shen2}, largely due to 
experimental uncertainties such as sample aging.  Despite this, our 
recent work on Bi2212 systems has found a straight Fermi 
surface segment that strongly resembles the Fermi surface of the 
stripe phase in La$_{1.28}$¥Nd$_{0.6}$¥Sr$_{0.12}$¥CuO$_{4}$¥  
\cite{Feng}.  This suggests that the charge stripe 
feature does exist in Bi2212, although the stripes are weaker as one 
can see the presence of quasiparticle like features near the d-wave 
node line.  The quasiparticle state is strongly suppressed in Nd-LSCO 
where static stripes exist.  

\subsection{Experimental Observation}
\label{Experimental Observation and Discussion2}
Fig.~\ref{Fig5ver5} shows ARPES data from pure and Zn doped samples.  
Again, two sets of data from Zn doped samples (Fig.~\ref{Fig5ver5}d-e) 
were presented to illustrate reproducibility.  Samples in 
Fig.~\ref{Fig5ver5}a-b are pure Bi2212 sample while sample in 
Fig.~\ref{Fig5ver5}c behaves as if it contains impurities.  The first 
qualitative observation is that the Zn doped samples show a stronger 
temperature dependence.  This temperature dependence correlates with 
the change of the spectral lineshape, with the Zn doped samples having 
a larger steplike background relative to the broad peak.  As the 
temperature is lowered, a sharp peak develops near 
$(0.8\pi,0)-(\pi,0)$, but without a dip structure 
(Fig.~\ref{TonyMbarver3}a).  The temperature induced change persists 
to very high energy.  This is more clearly visible in the spectral 
lineshape of the sample in Fig.~\ref{Fig5ver5}d.  When the peak moves 
away from E$_{f}$¥, the spectra for the two temperatures have 
different curvatures, with the 100K spectra being concave down and the 
20K spectra being concave up.  This behavior is beyond experimental uncertainties as 
we will discuss later.  The enhanced temperature dependence with Zn 
doping is consistent with the empirical result that the temperature 
dependence of ARPES gets stronger as the magnetic correlations get 
stronger.  In Sr$_{2}$¥CuO$_{2}$¥Cl$_{2}$¥ we see a much stronger 
temperature dependence on an energy scale of approximately 1eV 
\cite{Kim}.  Sr$_{2}$¥CuO$_{2}$¥Cl$_{2}$¥ is a Mott insulator with 
strong magnetic correlation.  In addition, the charge ordered 
manganite, Pr$_{0.5}$¥Sr$_{0.5}$¥MnO$_{3}$¥ shows a temperature 
dependence of spectral weight up to 1.2eV \cite{Chainani}.  To 
contrast, overdoped Bi2201 (the one plane version of Bi2212) with weak 
magnetic correlation does not show much temperature dependence (see 
next paragraph).  If Zn induces a local antiferromagnetic region 
around itself, it will presumably show a strong temperature 
dependence.  This lends itself to the idea that Zn strengthens the 
tendency of microscopic phase separation.

We now move to the second aspect of the temperature dependence of the 
ARPES spectra from the Zn doped samples.  While the spectral weight is 
gained near $(\pi,0)$, it is reduced near $(0.3\pi,0)-(0.5\pi,0)$.  At 
first glance, it appears there is a {\bf q} dependent spectral weight 
shift \cite{Shen2}.  The data from the two Zn doped samples are quite 
consistent with each other although they do differ in subtle ways.  
Judging from the sharpness of the feature, the sample in 
Fig.~\ref{Fig5ver5}e is more overdoped than sample in 
Fig.~\ref{Fig5ver5}d.  As shown before, ARPES lineshapes are extremely 
sensitive to doping changes \cite{Loeser} \cite{Ding}.  To check 
whether the apparatus worked appropriately, we show in 
Fig.~\ref{Fig6ver1} ARPES data recorded at 20K and 100K for two Bi2201 
samples taken under identical conditions as those of Bi2212 samples.  
As the temperature is lowered, the two sets of data match to within 
our error.  In our experimental runs we checked the Bi2201 many times; 
the fact that they matched so well in all cases gave us confidence 
that the experimental apparatus worked properly.  It is important to 
note that the calibration run using Bi2201 was not done across the 
superconducting transition so no significant lineshape change was 
involved and the situation was simpler.

While samples in Fig.~\ref{Fig5ver5}a-b do not show any change in 
n({\bf k}) with temperature, samples in Fig.~\ref{Fig5ver5}c,d,e show 
subtle changes.  Fig.~\ref{Znnofk} depicts the integrated spectral 
weight, n({\bf k})$=\int$A$(\omega$,{\bf k}) f($\omega$) d$\omega$ 
versus $\mid${\bf k}$\mid$ for the samples in 
Fig.~\ref{Fig5ver5}b,c,d.  Note that the n({\bf k}) curves have 
different shapes so the change cannot be simply corrected by a shift 
in {\bf k} (or a tilt in angle) meaning the difference cannot be trivially 
attributed to angular irreproducibility with temperature.  This point 
will be elaborated upon in detail later.  Taken naively, it appears 
that the spectral weight is shifted from {\bf k}$\sim$$(0.3\pi,0) - 
(0.5\pi,0)$ to $(0.8\pi,0) - (\pi,0)$ region with a $\mid${\bf 
Q}$\mid$ of about $0.4\pi - 0.5\pi$.  

\subsection{Discussion of Temperature Dependence Results}
Because the {\bf Q} value of $(0.4\pi,0)\sim(0.5\pi,0)$ is quite similar to what one 
expects from stripes, we attributed this {\bf q} shift to the stripe 
instability \cite{Shen2}.  We were further encouraged by the fact that 
the {\bf q} shift is more pronounced in Zn doped samples because Zn 
doping presumably enhanced the $\frac{1}{8}$ anomaly.  This is also 
consistent with the fact that the spectra along (0,0) to $(\pi,\pi)$ 
direction of Zn doped samples look like that of LSCO where the 
evidence for stripes is the strongest \cite{Tranquada} \cite{Koike}.  
However, there are some important points to be addressed within that 
interpretation.

One important issue is whether this effect of a change in n({\bf k}) 
was due to angular misalignment on our part.  It is emphatically 
stated that our results were reproducible and exhaustively studied.  
The following statements are included to emphasize resoundingly the 
veracity of the qualitative aspect of this observation if not the 
quantitative as well.  For the sake of argument, it is conjectured 
that the variation in n({\bf k}) is due to some angular mismatch in 
going from above T$_{c}$¥ to below T$_{c}$¥.  To explore this 
possibility, we utilized an ARPES system with finer momentum (angular) 
resolving power than previously used.  The difference is at least a 
factor of {\it seven} better than that for the typical ARPES system.  
The resulting data (taken in angle mode) is presented for a 
Bi$_{2}$¥Sr$_{2}$¥Ca(Cu$_{0.99}$¥Zn$_{0.01}$¥)$_{2}$¥O$_{8 + \delta}$¥ 
sample in Fig.~\ref{try02}.  Data is presented for both an aged and a 
fresh sample.  It is noted that no pair of hot and cold curves overlay 
each other, that is to say, assuming some kind of angular misalignment 
in cooling the apparatus, one would expect a given high temperature 
curve to have some match among the group of cooler curves.  That is 
not the case.  A realistic question would be if sample aging 
contributed to this effect.  It is noted that the intentionally aged 
sample also has no pair of matching curves.  Now, of course, thermal 
broadening must be taken into account as well as the fact that there 
is some amount of change due to the superconducting state {\it 
especially} near $(\pi,0)$.  However, the points chosen were not close 
to $(\pi,0)$ and thermal broadening of this magnitude does not account 
for the differences seen.  To investigate this even further, the 
spectral weight was integrated for both samples and plotted versus 
{\bf k} in Fig.~\ref{try01}.  The data from the fresh Zn sample 
reproduces the results of Fig.~\ref{Fig5ver5} and Fig.~\ref{Znnofk} 
and are consistent with our earlier results \cite{Shen2}.  The reader 
can readily discern that in the very least, the shapes of the n({\bf 
k}) do not match one another, either for the fresh or aged sample, no 
matter how they are translated in {\bf k} (an action correcting for 
the supposed angular misalignment).  Although the interpretation of 
the data may change in light of new findings or information, the 
experimental observation of this effect is without question.

An additional issue to be addressed concerns periodicity of the {\bf Q} 
structure.  A central premise in solid state physics is that 
periodicity seen in the first Brillouin zone at {\bf q} should also be 
seen at {\bf q}$+${\bf G} in the second Brillouin zone, where {\bf G} 
is the usual reciprocal lattice vector.  That is to say that the 
effect we found between $[0,\pi]$ should be realized between 
$[\pi,2\pi]$ as well.  This observation would confirm our estimate of 
$\mid${\bf Q}$\mid$ and establish a true periodicity.  However much 
this is desired, the experiment, nonetheless, dictates its 
impracticality.  This would involve the sampling of more points, which 
would increase the likelihood of aging effects, which have been so 
detrimental to the experimental act.  However, in an attempt to 
satisfy a probable and fair question on the part of the reader, we 
utilized our system at ALS with the higher photon flux and sampling 
density.  Our results on a Zn doped sample are shown in 
Fig.~\ref{2ndbz} as an n({\bf k}) plot extending into the second 
Brillouin zone.  What this plot shows is the suppressed spectral weight 
between $[\pi,2\pi]$.  This is most likely due to matrix element 
effects \cite{Randeria}.  The error bars in the second zone comprise a 
larger percentage of the total as compared to the first zone.  Taking 
a difference of these above and below T$_{c}$¥ would be less 
informative than doing so in the first zone, and we are already 
pushing the experimental limitations of our apparatus.  It is possible 
to revisit this question.  

It has been recently been suggested that n({\bf k}) should be 
temperature {\em independent}.  Further, all temperature induced 
changes are confined in the window of $3\Delta$ with $\Delta$ being 
the superconducting gap \cite{Randeria}.  Following that assumption, 
any temperature dependence ought to be attributable to experimental 
artifact.  However, after careful consideration of some key results, 
we find the assertion of spectral weight conservation to be 
inconsistent.  Fig.~\ref{Rander3b} reproduces data \cite{Randeria} 
published by the same authors cited in ref.~\ref{Campuzano} as a proof 
for the lack of temperature dependence \cite{Randeria}.  In 
ref.~\ref{Randeria}, in distinct contrast to ref.~\ref{Campuzano}, the 
same authors state that spectral weight {\em is} conserved for {\bf 
k}={\bf k}$_{f}$¥ but {\em is not} conserved for {\bf k}$\neq${\bf 
k}$_{f}$¥ (which is the region of relevance).  After subtracting out 
temperature dependent background, the same authors claimed a 10$\%$ 
change in spectral weight between T=13K and T=105K. While the reason 
for this inconsistency of claims by the same authors needs to be 
investigated, it is clear that the claim of spectral weight 
conservation has not been established contrary to 
ref.~\ref{Campuzano}.

Some other points are in order concerning the issue of angular 
misalignment \cite{Campuzano}.  The n({\bf k}) curves (see 
Fig.~\ref{Znnofk} and Fig.~\ref{try01}) show shape changes with 
temperature that cannot be explained by a rigid angular shift, as 
discussed earlier.  In other words, one cannot slide the two n({\bf 
k}) curves at different temperatures to match each other because they 
have different shapes, contrary to the example of 
ref.~\ref{Campuzano}.  As shown by the results of Fig.~\ref{Fig6ver1}, 
the experimental setup works appropriately.  The fact that the spectra 
at two temperatures match so well in the Bi2201 samples in all cases 
rule out the possibility of angle tilt of our sample manipulator as 
the temperature changed.  Any possible angle shift can only stem from 
the mechanical flatness of the sample surface which is a random 
quantity.  However, we have observed a strong correlation between the 
{\bf q} shift and other electronic structure changes: the destruction 
of the normal state quasiparticle along the $(\pi,\pi)$ direction, the 
suppression of the superconducting gap, the broadening of the 
superconducting peak, and the suppression of the superconducting dip 
near $(\pi,0)$.  While the surface flatness can introduce error, this 
random variable alone cannot explain the {\em systematic} changes 
observed.

To further test the issue of charge stripes in Bi2212, we need to find 
a better way to passivate the surface.  Although Bi2212 has a very 
stable surface in comparison to most cuprate samples, it appears that 
it is not stable enough if one wants to detect very subtle effects on 
the scale of a few percent unless extreme care is taken with respect 
to vacuum quality.  This is the main source of uncertainty of our 
previous result \cite{Shen2}.  Rather than wrestle further with this 
uncertainty, perhaps an alternative methodology would serve.  To test 
the idea of stripes, we need to study materials where the evidence for 
stripes from other experiments is unambiguous (this will establish 
what one should expect from stripes, although some theoretical ideas 
have already been proposed).  This has been done recently in the 
stripe phase of La$_{1.28}$¥Nd$_{0.6}$¥Sr$_{0.12}$¥CuO$_{4}$¥ 
\cite{Zhou}.  The effect is very strong so a modulation of n({\bf k}) 
with 0.5$\pi$ periodicity is seen without the need to do the 
difference curve.  More recently, we have used an alternative approach 
to the problem by looking at the Fermi surface nesting feature which 
is strongly present in the stripe phase \cite{Zhou}.  With a global 
mapping in n({\bf k}), we have found clear Fermi surface nesting 
features similar to those seen in the static charge order phase 
\cite{Feng}.  This does support the notion of the presence of stripes 
even in Bi2212 when one looks at the data with high frequency.  
However, the stripe instability is weaker in Bi2212 than in the LSCO 
system.  In Bi2212, one can still see a quasiparticle like feature 
along $(\pi,\pi)$ direction while this is totally absent in LSCO.

\section{Summary} 
\label{Summary2}
In conclusion, we recast the most important qualitative observation of 
our experiment - the destruction of the quasiparticle peak along 
$\Gamma$Y with relatively small amounts of Zn impurities.  Whatever 
the microscopic mechanism causing the change, be it simple impurity 
scattering or some random pinning of the phase separated domains, this 
observation is significant to the understanding of the relevance of 
quasiparticles in the normal state.  Data in Fig.~\ref{Fig3ver5}a,b,e,f 
may reasonably be interpreted within the context of the quasiparticle 
picture, a concept that is the foundation of the modern theory of 
solids and is extensively used to address the high T$_{c}$¥ problem.  
Yet, one would be very hard pressed to call the data in 
Fig.~\ref{Fig3ver5}c-d as reflecting quasiparticles, and 
superconductivity with T$_{c}$¥ as high as 83K survived.  Taken 
naively, the existence of conventional quasiparticles with well 
defined {\bf k} does not seem to be necessary for the realization of 
high temperature superconductivity.  We also have studied temperature dependence in the ARPES data from the  
Bi$_{2}$¥Sr$_{2}$¥Ca(Cu$_{1-x}$¥Zn$_{x}$¥)$_{2}$¥O$_{8+\delta}$¥ 
system.  With  Zn doping, we observed strong 
temperature dependence in the data which persists up to very high 
energy.  This observation is consistent with the idea that Zn 
enhances the tendency of electronic inhomogeneity, a 
result consistent with NMR result \cite{Alloul}.  

We acknowledge S.A. Kellar, X.J. Zhou, and P. Bogdanov for technical 
help, and R.B. Laughlin, D. S. Dessau, A.J. Millis, M. 
Norman, B.O. Wells, S.A. Kivelson, A. Fujimori, D. J. Scalapino, H. 
Eisaki, N. Nagosa, and D. van der Marel for helpful discussions.  
ARPES experiments were performed at SSRL which is operated by the DOE 
Office of Basic Energy Science, Division of Chemical Sciences.  The 
Office's Division of Material Science provided support for this 
research.

\newpage

\begin{figure}
\caption{ARPES data along (0,0) to $(\pi,\pi)$ cut for several types 
of Bi$_{2}$¥Sr$_{2}¥$CaCu$_{2}$¥O$_{8+\delta}$¥.  The number near the 
spectrum indicates the {\bf k}-space point.  (a) and (b) are pure with 
T$_{c}$¥ of 91K; (c) and (d) are Zn doped with T$_{c}$¥ of 83K; (e) is 
10\% Dy doped with T$_{c}$¥ of 65K (underdoped sample); (f) is
an oxygen reduced (underdoped) with T$_{c}$¥ of 67K. All spectra were 
collected at 100K under comparable conditions.  For the Zn doped 
sample (d), spectra recorded below T$_{c}$¥ (gray line) is the same 
within the experimental uncertainty.  The topmost spectrum in panel 
(a) is the {\it average} of the others (see text for discussion).}
\label{Fig3ver5}
\end{figure}

\begin{figure}
\caption{ARPES spectra of Bi$_{2}$¥Sr$_{2}¥$CaCu$_{2}$¥O$_{8+\delta}$¥ 
recorded at $(\pi,0)$ for (a) pure (T$_{c}$¥$\approx$91K) and Zn doped 
(T$_{c}$¥$\approx$83K) in the normal (100K) and superconducting (20K) 
states and for (b) pure underdoped and overdoped in the superconducting state (20K).}
\label{TonyMbarver3}
\end{figure}

\begin{figure}
\caption{ARPES data along (0,0) to $(\pi,0)$ cut for 
Bi$_{2}$¥Sr$_{2}¥$CaCu$_{2}$¥O$_{8+\delta}$¥ for (a) pure 
(T$_{c}¥\approx$89K); (b) pure (T$_{c}¥\approx$91K); (c) pure 
(T$_{c}¥\approx$88K); (d) Zn doped (T$_{c}¥\approx$78K); (e) Zn doped 
(T$_{c}¥\approx$83K). 100K data are represented by the gray lines while 
20K data are represented by the black lines.}
\label{Fig5ver5}
\end{figure}

\begin{figure}
\caption{ARPES data along $\Gamma\bar{M}$ cut for two samples (a and 
b) of the overdoped version of the one plane compound, 
Bi$_{2}$¥Sr$_{2}$¥CuO$_{6}$¥, with T$_{c}¥\approx$8K. 100K data are 
represented by the gray lines while 20K data are represented by the 
black lines.}
\label{Fig6ver1}
\end{figure}

\begin{figure}
\caption{Bi$_{2}$¥Sr$_{2}¥$CaCu$_{2}$¥O$_{8+\delta}$¥ n({\bf k}) plots 
versus $\mid${\bf k}$\mid$ for (0,0) to $(\pi,0)$ for (a) pure 
(T$_{c}¥\approx$91K, Fig.~\ref{Fig5ver5}b); (b) pure 
(T$_{c}¥\approx$88K, Fig.~\ref{Fig5ver5}c); and (c) Zn doped 
(T$_{c}¥\approx$78K, Fig.~\ref{Fig5ver5}d).  Open squares represent 
100K data; filled circles represent 20K data.  Error bars of $\pm$1\% 
of the total are not included because of scale.}
\label{Znnofk}
\end{figure}

\begin{figure}[t]
\caption{Data from 
Bi$_{2}$¥Sr$_{2}$¥Ca(Cu$_{0.99}$¥Zn$_{0.01}$¥)$_{2}$¥O$_{8 + \delta}$¥ 
along (0,0) to $(\pi,0)$ cut for T=100K(solid), 20K(gray) for (a) 
fresh sample and (b) aged sample.  The {\bf k} is indicated by the 
position along (0,0) to $(\pi,0)$ in units of $\frac{\pi}{a}$ given by 
the number to the right for the topmost and bottom spectra.  
Successive pairs of spectra are evenly spaced to show detailed 
evolution of the features.  Data are taken in angle mode.}
\label{try02}
\end{figure}

\begin{figure}
\caption{(a) n({\bf k}) plots along $\Gamma\bar{M}$ for fresh surface 
data (Fig.~\ref{try02}a).  (b) n({\bf k}) from aged surface data 
(Fig.~\ref{try02}b).}
\label{try01}
\end{figure}

\begin{figure}
\caption{n({\bf k}) along $\Gamma\bar{M}$ for 
Bi$_{2}$¥Sr$_{2}$¥Ca(Cu$_{0.99}$¥Zn$_{0.01}$¥)$_{2}$¥O$_{8 + \delta}$¥ 
extending into the second Brillouin zone taken in the normal state 
(T=100K).}
\label{2ndbz}
\end{figure}

\begin{figure}
\caption{Data taken from ref.~\ref{Randeria} showing near optimal 
Bi$_{2}$¥Sr$_{2}¥$CaCu$_{2}$¥O$_{8+\delta}$¥ (T$_{c}$¥$\approx$87) 
spectra near the $\bar{M}$ point above (105K) and below T$_{c}$¥ 
(13K).}
\label{Rander3b}
\end{figure}


\begin{references}
\bibitem{comment2} The low temperature data does allow the 
possibility of a gap, but more needs to be done before a conclusion 
can be drawn.
\bibitem{comment3} Empirically, one may naively think that the peak 
and dip minimum energy actually scale with each other.  This is 
reasonable as the dip minimum energy is basically determined by the 
peak energy and half of the width.
\end{references}
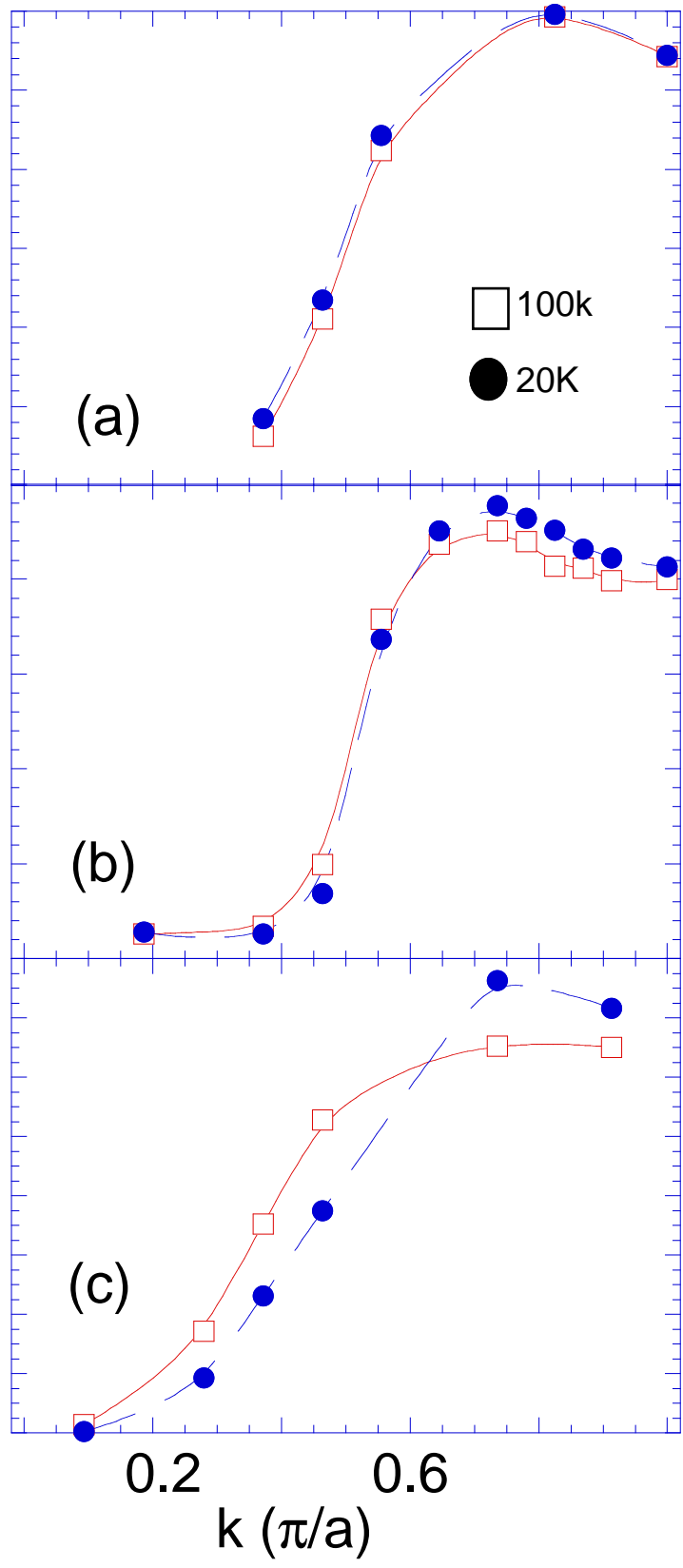
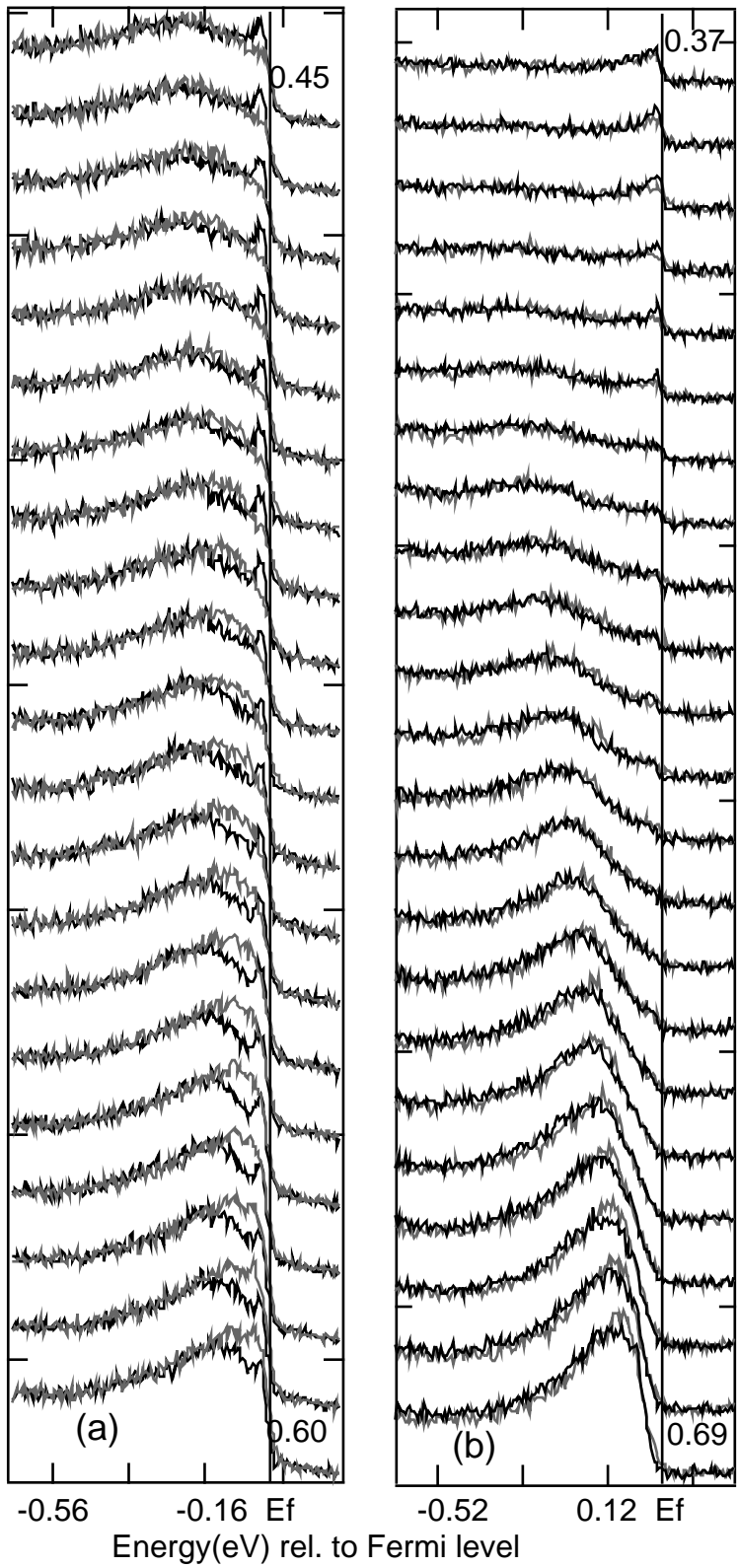
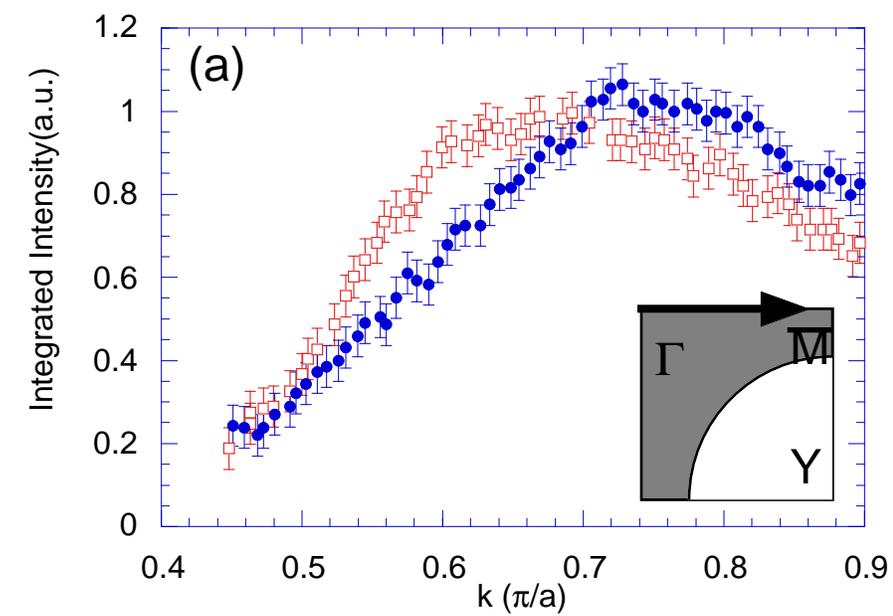
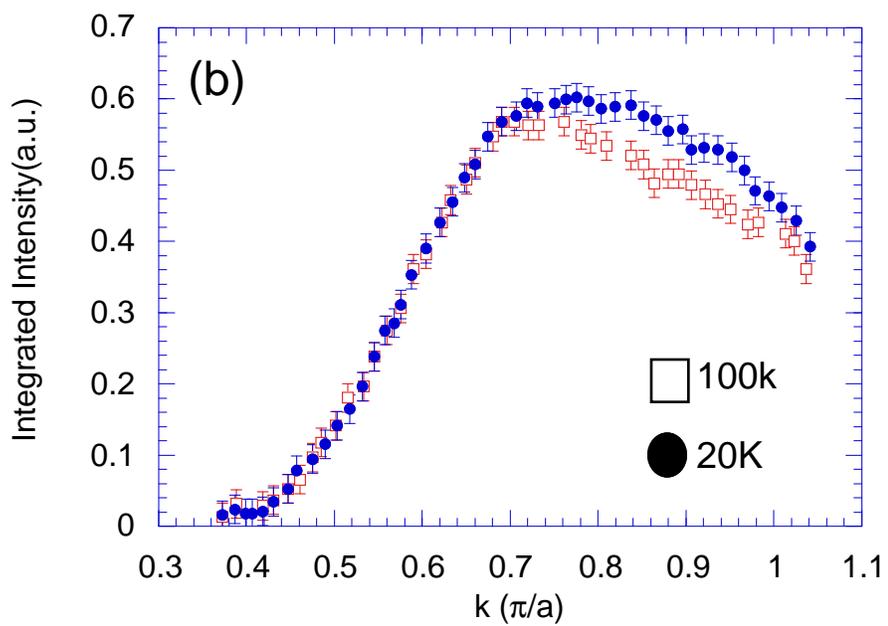
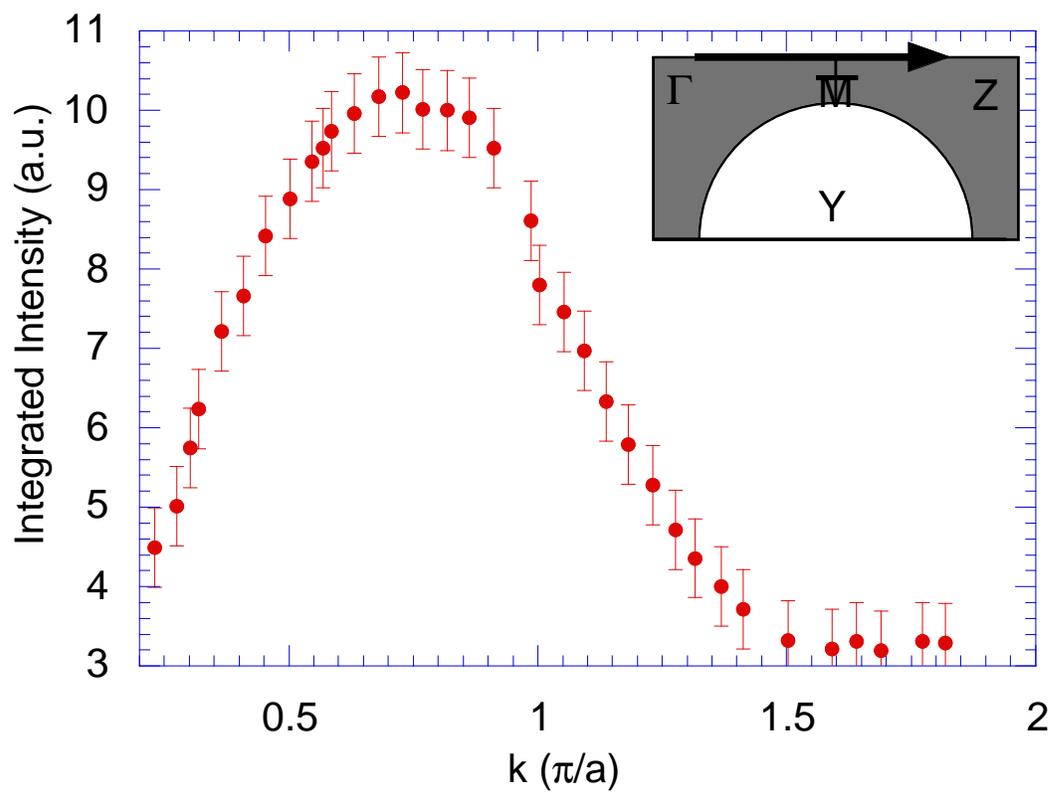
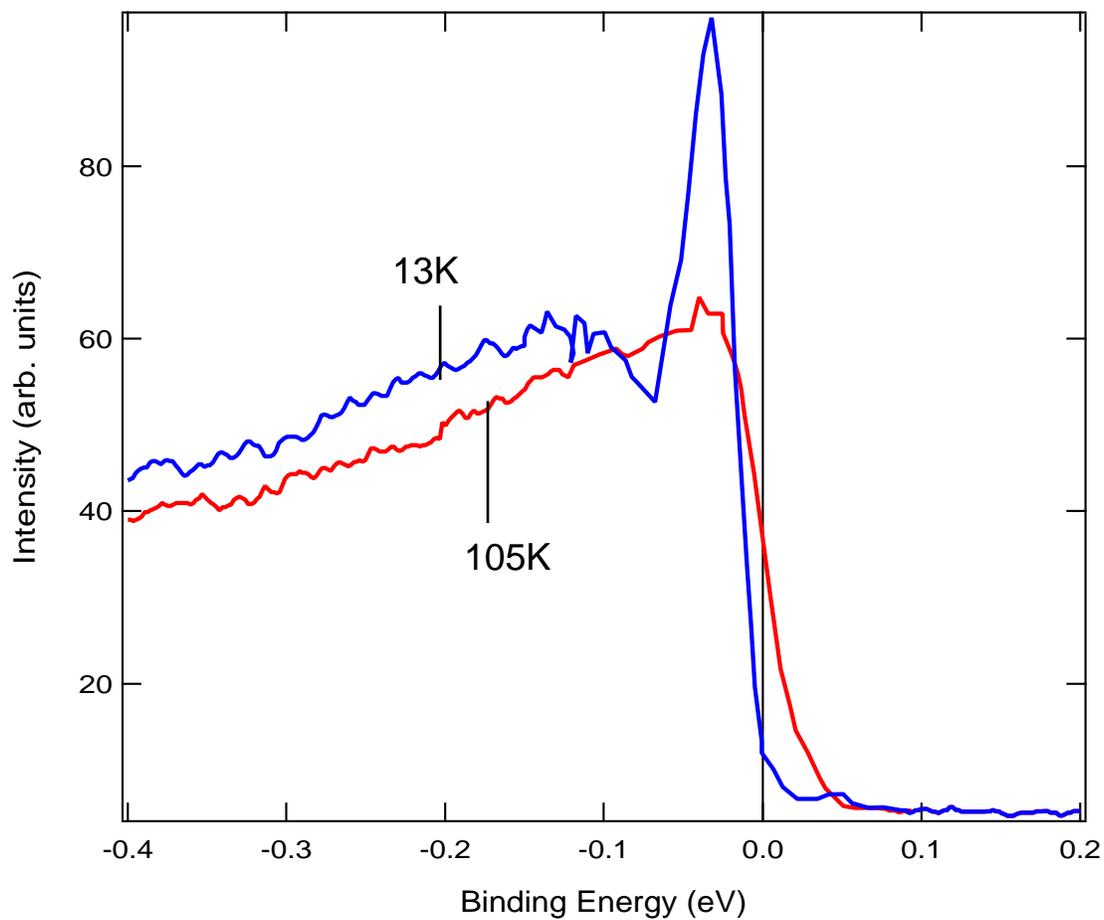
\end{document}